\documentclass[%
 reprint,
 amsmath,amssymb,
 aps,
 prb,
]{revtex4-2}

\usepackage{graphicx}
\usepackage{dcolumn}
\usepackage{bm}
\usepackage{hyperref}

\begin{document}

\preprint{}

\title{Electronic Properties and Interlayer Interactions in Antimony Oxide Homo- and Heterobilayers}

\author{Stefan Wolff}
\email{stefan.wolff@fau.de}
\affiliation
{Department of Physics, Chair of Experimental Physics, Friedrich-Alexander-Universit\"at Erlangen-N\"urnberg (FAU), Staudtstr. 7, 91058 Erlangen, Germany}

\author{Roland Gillen}
\affiliation
{Department of Physics, Chair of Experimental Physics, Friedrich-Alexander-Universit\"at Erlangen-N\"urnberg (FAU), Staudtstr. 7, 91058 Erlangen, Germany}

\author{Janina Maultzsch}
\affiliation
{Department of Physics, Chair of Experimental Physics, Friedrich-Alexander-Universit\"at Erlangen-N\"urnberg (FAU), Staudtstr. 7, 91058 Erlangen, Germany}

\date{\today}

\begin{abstract}
Antimony shows promise as a two-dimensional (2D) mono-elemental crystal, referred to as antimonene. When exposed to ambient conditions, antimonene layers react with oxygen, forming new crystal structures, leading significant changes in electronic properties. These changes are influenced by the degree of oxidation. Utilizing Density Functional Theory (DFT) calculations, stable configurations of bilayer antimony oxide and their corresponding electronic properties are studied. Additionally, different stacking arrangements and their effects on the physical properties of the materials are investigated. Furthermore, the analysis encompasses strain-free hetero-bilayers containing both pristine and oxidized antimonene layers, aiming to understand the interplay between these materials and their collective impact on the bilayer properties. Our results provide insight into how the properties of antimony-based bilayer structures can be modified by adjusting stoichiometry and stacking configurations.
\end{abstract}

\keywords{antimonene, 2D materials, tunable bandgap, density functional theory (DFT)}

\maketitle

\section{Introduction}
In recent years, two-dimensional (2D) materials have captivated the scientific community due to their extraordinary electronic and mechanical properties, opening new possibilities for various technological applications. Among these, the group-15 elements, also known as pnictogens, are very promising candidates for several different applications. One of these elements, which is considered to have great potential, is antimony. As a 2D material it forms a buckled honeycomb lattice in its most stable configuration, the $\mathrm{\beta}$-phase, which is called antimonene.~\cite{Lloret2019,Ares2016,Gibaja2016,Ares2018,Zhang2016,Zhang2018,Pumera2017} Its electronic band gap of about 2.4\,eV~\cite{Zhang2015,Akturk2015,Wang2015,Wang2017} makes this material a great candidate for electronic and optoelectronic devices.~\cite{Wang2017,Pizzi2016,Chen2018,Zhang2019,Wang2018,Wang2019} Additionally, antimonene has also proven to be well suited for catalysis~\cite{Lloret2019,Li2017}, as a material for energy storage~\cite{Gu2017,Martinez-Perinan2018}, and for sensing,~\cite{Mayorga-Martinez2019} and biomedical applications such as cancer therapy.~\cite{Tao2017,Carrasco2023,Lucherelli2023} Antimonene exhibits, in contrast to many other 2D materials, residual covalent interlayer bonds, which can have a high impact on the electronic properties of few-layer structures, resulting in a significant decrease of the electronic band gap, with the inclusion of additional layers.~\cite{Zhang2015,Akturk2015}

However, despite its promising potential, the practical application of antimonene is hindered by its reactivity with oxygen under ambient conditions.~\cite{Lloret2019,Ares2016,Gibaja2016,Ares2018,Abellan2017,Ji2016,Kistanov2018,Wolff2020,Assebban2020} The interaction with oxygen leads to the formation of antimony oxide layers, introducing new crystal structures with distinct electronic and vibrational properties. Even though the exact configurations of these crystals are not entirely clear yet, we recently suggested novel 2D antimony oxide arrangements.~\cite{Wolff2020,Assebban2020} If a controlled oxidation process can be achieved it may be possible to tailor the electronic band gap to values between 2.0\,eV and 4.9\,eV.~\cite{Wolff2020,Assebban2020} The degree of oxidation becomes a critical factor for the resulting electronic properties, making it essential to explore and understand the stable configurations of bilayer antimony oxide and their corresponding electronic behaviors. Furthermore, an investigation of the interactions of the 2D antimony oxide crystals with neighboring materials is of particular interest, especially considering potential applications.

In this study, we employ DFT calculations to explore a set of stable configurations of bilayer antimony oxide and investigate their electronic properties. We analyze different stacking arrangements of the antimony oxide layers and determine their influence on the characteristics of the material.

Our previous investigations of bi- and trilayer antimonene have shown that oxidation occurs on the surface layers of the materials, thus resulting in a predetermined stacking configuration of the structures. In the bilayer case, both layers are oxidized, in contrast to the trilayer structures which consist of a non oxidized antimonene layer sandwiched between two antimony oxide layers.~\cite{Wolff2020} Due to the lattice mismatch of pristine antimonene and the oxidized antimony oxide layers, the trilayer structures exhibit a high amount of strain which is absent for the bilayer structures. Therefore we herein focus on the oxidized antimony bilayers which might be more realistic for real world applications.

\begin{figure}[t]
    \centering
    \includegraphics[width=\linewidth]{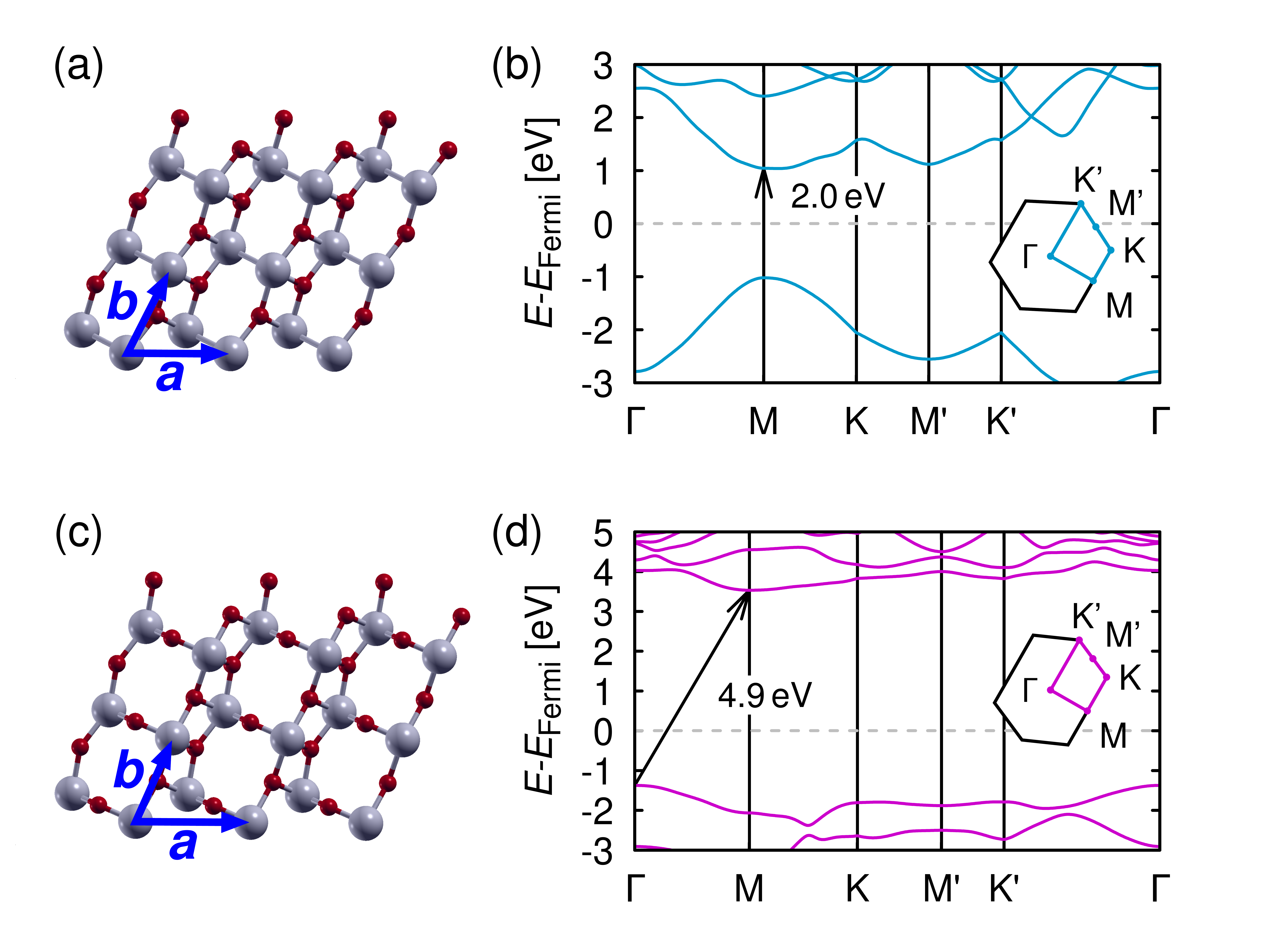}
    \caption{Single layers of the Sb$_2$O$_2$ structure in (a) and the Sb$_2$O$_3$ structure in (c) with the corresponding monolayer band structures and the respective band gaps of 2.0\,eV and 4.9\,eV in (b) and (d), see also Ref.~\cite{Wolff2020}. Both structures are shown from the top. Oxygen atoms are displayed in red and antimony atoms in gray. The vectors $\bm{a}$ and $\bm{b}$ span the unit cell of the respective monolayer. The chains of alternating oxygen and antimony atoms are oriented along the $\bm{b}$ direction. For clarity, only the top layers of the bilayer structures are displayed here. The fully relaxed stacking arrangements of the respective bilayer structures with different stacking configurations are displayed in Figure~\ref{fig:SbO_2L_structures}.}
    \label{fig:monolayer}
\end{figure}

We create antimony oxide bilayers from monolayers with different stacking configurations. In our previous work, we found two different stoichiometric configurations for 2D antimony oxide to be stable: The monolayer unit cell of the first structure consists of two antimony atoms and two oxygen atoms, i.e. Sb$_2$O$_2$. The other stable monolayer configuration comprises two antimony atoms and three oxygen atoms, i.e. Sb$_2$O$_3$.~\cite{Wolff2020} Both configurations show very similar structural features, which are chains of alternating oxygen and antimony atoms. In the case of the Sb$_2$O$_2$ structure, these chains are connected via bonds between two antimony atoms from neighboring chains, whereas in the Sb$_2$O$_3$ structure, the bonds connecting the neighboring chains are replaced by the additional oxygen atom.~\cite{Wolff2020} Figure~\ref{fig:monolayer}(a) and (c) show the top layer of the initial Sb$_2$O$_2$ bilayer and the initial Sb$_2$O$_3$ bilayer, respectively. The corresponding electronic band structures of these monolayers are shown in Figure~\ref{fig:monolayer}(b) and (d), which were previously calculated in Ref.~\cite{Wolff2020}.

\section{Results and Discussion}

\subsection{Homobilayer Stacking Configurations}

\begin{figure}[t]
    \centering
    \includegraphics[width=\linewidth]{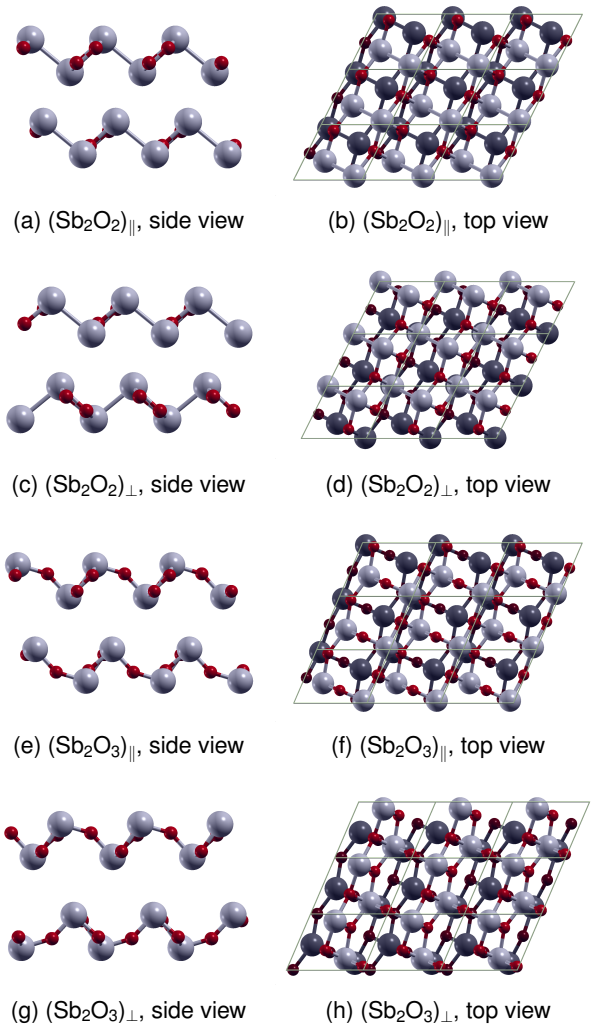}
    \caption{The different investigated stacking configurations (a), (b) (Sb$_2$O$_2$)$_\parallel$, (c), (d) (Sb$_2$O$_2$)$_\perp$, (e), (f) (Sb$_2$O$_3$)$_\parallel$, (g), (h) (Sb$_2$O$_3$)$_\perp$ shown from the side and from the top. The side view of the structures is aligned such that the chains of alternating antimony (gray) and oxygen (red) atoms extend into the drawing layer. This means the lattice vector $\bm{b}$ (the direction of the chains, see Figure~\ref{fig:monolayer}) is perpendicular to the paper plane for the side view. In the top view images, the atoms in the lower layer are colored in a darker hue.}
    \label{fig:SbO_2L_structures}
\end{figure}

In Figure~\ref{fig:SbO_2L_structures}(a), (b) and (e), (f), we show the original stacking order, which is derived from oxidizing a naturally stacked antimonene bilayer. There the chains of oxygen and antimony atoms are all parallel to each other. These structures are referred to as (Sb$_2$O$_2$)$_\parallel$ and (Sb$_2$O$_3$)$_\parallel$. In addition, our calculations show that different stacking configurations may exist, in which the chains of neighboring layers are perpendicular to each other. Structures with this configuration are denoted by (Sb$_2$O$_2$)$_\perp$ and (Sb$_2$O$_3$)$_\perp$, see Figure~\ref{fig:SbO_2L_structures}. The difference between the length of the lattice vectors between the respective mono- and bilayer structures for the Sb$_2$O$_2$ structures is less than 0.5\,\% and the bond lengths differ by less than 0.3\,\%. These variations are even smaller for the Sb$_2$O$_3$ structures. We assume that all of the structures presented in Figure~\ref{fig:SbO_2L_structures} are stable, since the difference in total energy per atom is on the order of a few meV. The difference in stoichiometry has been accounted for by calculating and subtracting or adding the total energy of the additional atoms respectively.

To further prove the stability of the newly constructed bilayer antimony oxide structures, in which the orientation of the Sb-O-chains of neighboring layers is perpendicular to each other, an additional translation of one layer with respect to the other is applied. In the case of the (Sb$_2$O$_2$)$_\perp$ structure, both initial configurations result in the same stacking order after relaxation. The relaxed configurations of the these structures are shown in Figure~\ref{fig:SbO_2L_structures}. The initial configurations can be seen in the Supporting Information (SI) in Figures~S1 and S2. For the shifted (Sb$_2$O$_3$)$_\perp$ configuration, a displacement along the direction of the Sb-O-chains was applied which did not relax back into the same structure, see SI Figure~S3. However, the difference of the total energy of both configurations is almost identical and the change of the charge density and the electronic band structure exhibits only minimal differences. The charge density and band structure of the shifted (Sb$_2$O$_3$)$_\perp$ configuration is shown in the SI in Figure~S4.

\subsection{Electronic Properties of Homobilayer Structures}\label{sec:SbSb}
\begin{figure}[t]
    \centering
    \includegraphics[width=\linewidth]{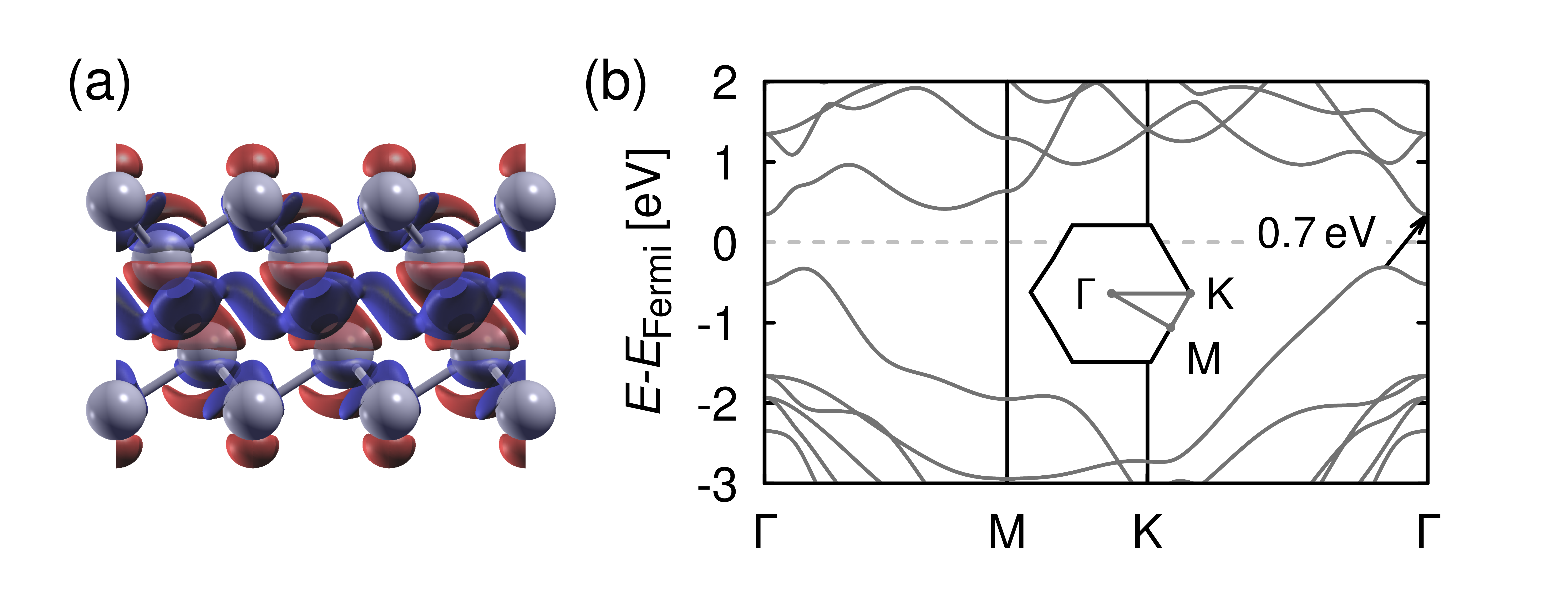}
    \caption{(a) redistribution of the positive (negative) charge density of bilayer antimonene arising from the formation of a bilayer as red (blue) clouds. The superposition of atomic densities and the charge density of the individual isolated layers is subtracted. (b) the corresponding electronic band structure obtained via HSE12 hybrid functionals.~\cite{hse12} The Fermi level is at zero of energy. The indirect electronic band gap has a value of about 0.7\,eV.}
    \label{fig:Sb_2L}
\end{figure}

The partially covalent bonding character of non-oxidized antimonene is well reflected by the significant overlap of the electron density between the layers, see Figure~\ref{fig:Sb_2L}(a). The red (blue) clouds represent the positive (negative) charge density redistribution arising from the formation of a bilayer. We subtracted the superposition of atomic densities and the charge density of the individual isolated layers from the bilayer charge density to emphasize the part of the charge density that stems form the formation of a bilayer. We will refer to this as ``interlayer charge density". The overlap of the electronic orbitals from neighboring layers leads to a significant reduction of the electronic band gap to a value of just below 0.7\,eV in our calculations using HSE12 hybrid functionals~\cite{hse12}, see Figure~\ref{fig:Sb_2L}(b). Other calculations even predict a semi-metallic character of few-layer antimonene.~\cite{Zhang2015,Akturk2015}

\begin{figure}[t]
    \centering
    \includegraphics[width=\linewidth]{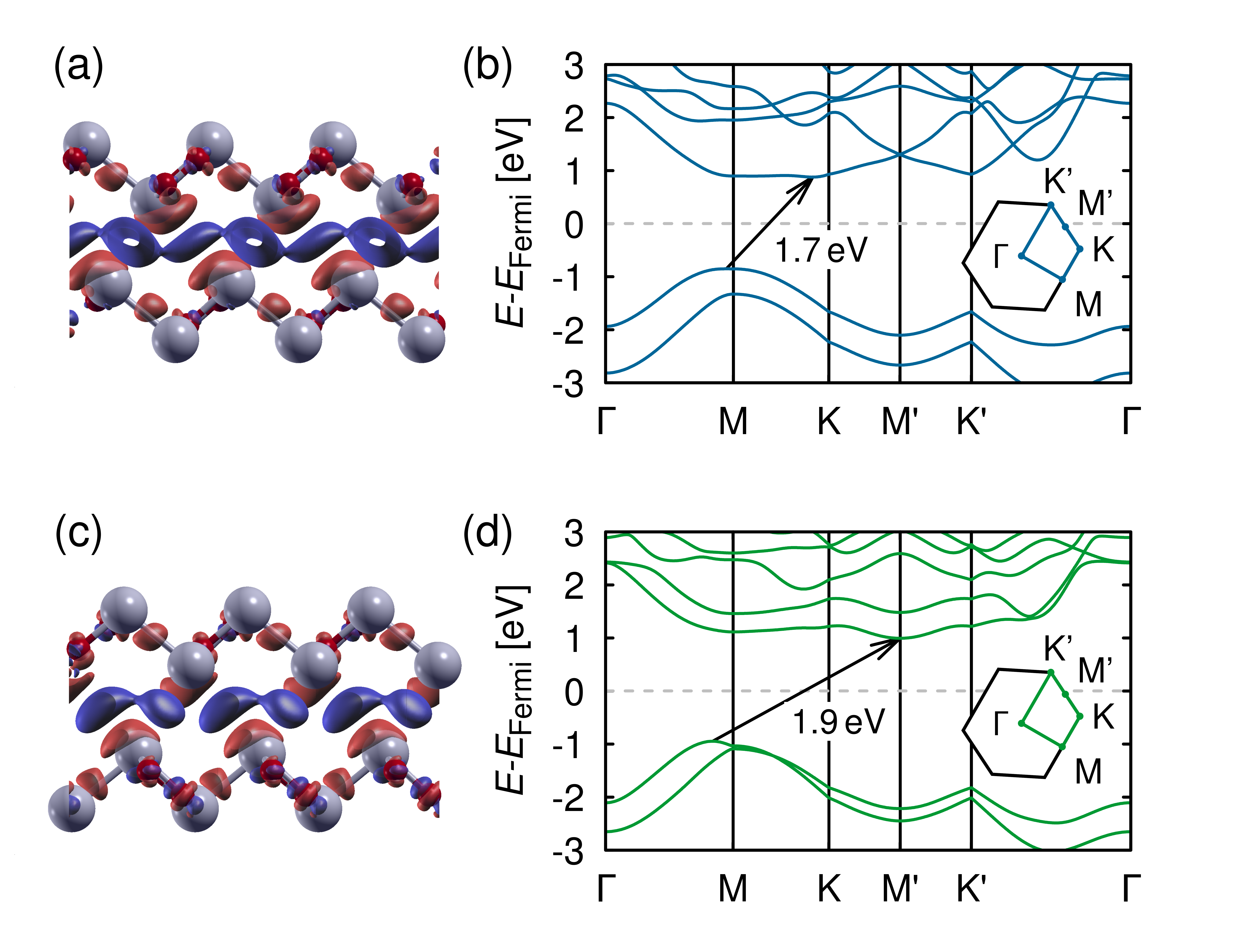}
    \caption{Bilayer charge density, after subtracting the superposition of atomic densities and the charge density of the individual isolated layers of bilayer (Sb$_2$O$_2$)$_\parallel$ and (Sb$_2$O$_2$)$_\perp$ in (a) and (c), respectively. The corresponding electronic band structures obtained via HSE12 hybrid functionals~\cite{hse12} are shown in (b) and (d). The Fermi level is set to the zero of energy. (Sb$_2$O$_2$)$_\parallel$ exhibits an electronic band gap of about 1.7\,eV; (Sb$_2$O$_2$)$_\perp$ has a slightly higher electronic band gap of about 1.9\,eV.}
    \label{fig:Sb2O2_2L}
\end{figure}

In contrast to the pristine antimonene bilayer structure, the overlap of the charge density of the (Sb$_2$O$_2$)$_\parallel$ configuration is reduced. However, as seen in Figure~\ref{fig:Sb2O2_2L}(a), an overlap of the charge density between the layers still persists. Despite this overlap, the electronic band structure exhibits an indirect band gap of about 1.7\,eV which is slightly lower than the monolayer band gap of about 2.0\,eV for this configuration.~\cite{Wolff2020} The valence band maximum (VBM) is close to the $M$ point along the ${\Gamma}M$ direction, whereas the conduction band minimum (CBM) is along the $MK$ line close to the $K$ point. The corresponding electronic band structure is depicted in Figure~\ref{fig:Sb2O2_2L}(b).

By changing the stacking configuration of the (Sb$_2$O$_2$)$_\parallel$ to the (Sb$_2$O$_2$)$_\perp$ structure, the overlap of the charge density between the layers is further reduced, see Figure~\ref{fig:Sb2O2_2L}(c). However, the electronic band gap is not changing drastically. The band gap has a value of 1.9\,eV [Figure~\ref{fig:Sb2O2_2L}(d)] and is very close to the electronic band gap of the respective monolayer structure. The VBM is still close to the $M$ point along the ${\Gamma}M$ line, but the CBM shifts significantly to the $M'$ point.

The overlap of the charge density thus depends on the stacking configuration, which also has an impact on the electronic properties of the material. The area where the charge densities intersect also appears to be a reliable indicator of the interactions between layers: In bilayer antimonene, the large overlap of the electron density caused by lone pair $d$ orbitals of neighboring layers suggests high interlayer interactions and leads to a small electronic band gap. By introducing oxygen into the structure, less orbitals might be available for the interlayer bonding, which can be seen by the reduction of the electron density between the two layers. However, interlayer charge density is maintained, thus still leading to a small reduction of the electronic band gap compared to monolayer Sb$_2$O$_2$. Alteration of the stacking configuration reduces the overlap of the electronic orbitals between the layers even further, such that both layers might be viewed as almost isolated, resulting in a band gap very close to the monolayer value.

\begin{figure}[t]
    \centering
    \includegraphics[width=\linewidth]{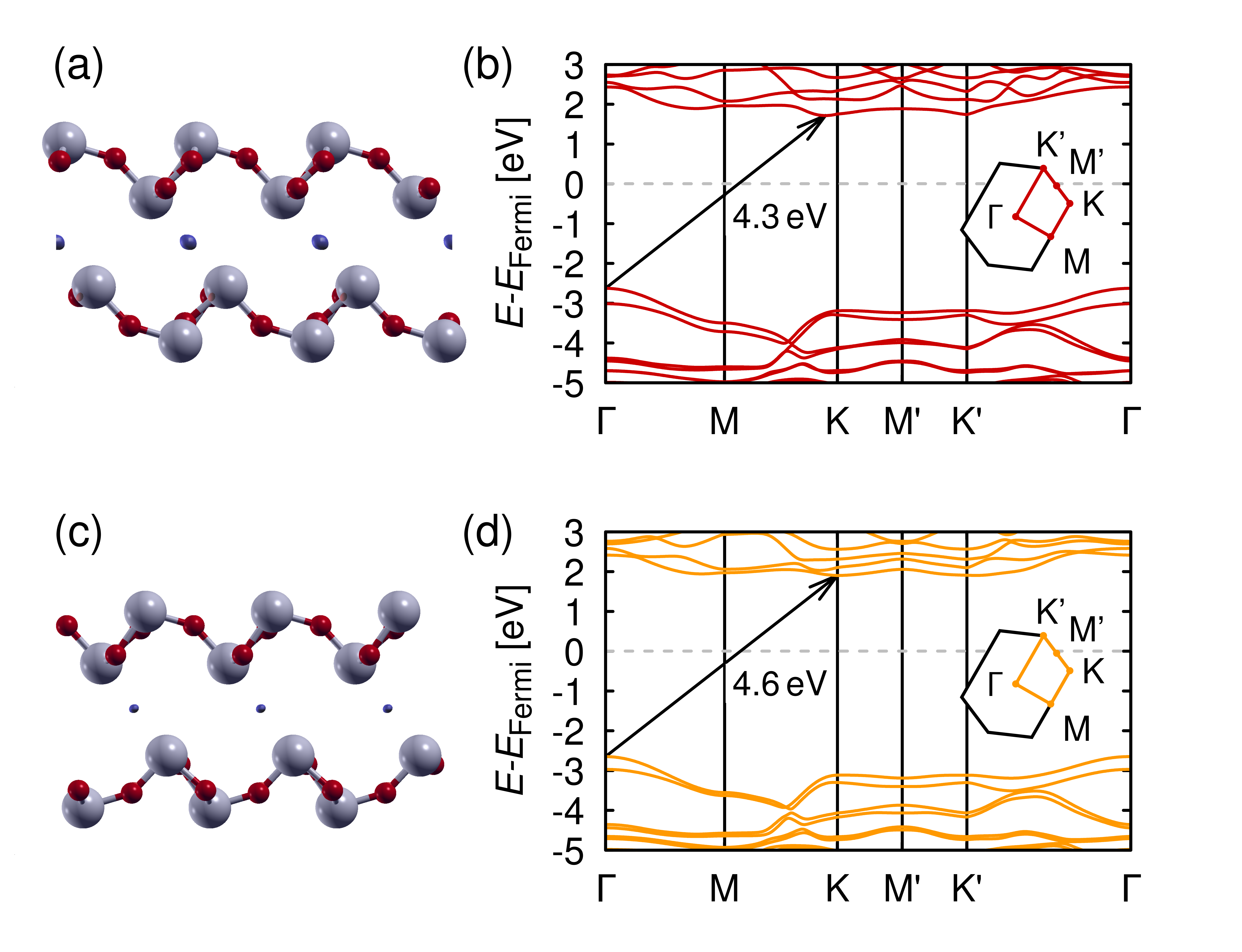}
    \caption{Bilayer charge density, after subtracting the superposition of atomic densities and the charge density of the individual isolated layers of bilayer (Sb$_2$O$_3$)$_\parallel$ (a), and (Sb$_2$O$_3$)$_\perp$ (c). The electronic band structures obtained via HSE12 hybrid functionals~\cite{hse12} for (Sb$_2$O$_3$)$_\parallel$ are illustrated in (b) for (Sb$_2$O$_3$)$_\perp$ in (d). The zero energy point aligns with the Fermi level. The electronic band gaps are 4.3\,eV and 4.6\,eV, respectively.}
    \label{fig:Sb2O3_2L}
\end{figure}

In the (Sb$_2$O$_3$)$_\parallel$ structure, the overlap of the orbitals of neighboring layers has vanished, see Figure~\ref{fig:Sb2O3_2L}(a), which would suggest a drastic decrease of the covalent interactions between the layers. In the monolayer case, for this configuration the band gap increases strongly to about 4.9\,eV with the addition of the additional oxygen atom.~\cite{Wolff2020} The electronic band structure of bilayer (Sb$_2$O$_3$)$_\parallel$ can be seen in Figure~\ref{fig:Sb2O3_2L}(b). In analogy to the (Sb$_2$O$_2$)$_\parallel$ structure, the electronic band gap is lower than the electronic band gap of the respective monolayer: The bilayer has an indirect electronic band gap of about 4.3\,eV. For this structure, the VBM is at the $\Gamma$ point of the Brillouin zone. Similar to the (Sb$_2$O$_2$)$_\parallel$ structure, the CBM is located close to the $K$ point.

The interlayer charge density of the (Sb$_2$O$_3$)$_\perp$ structure, see Figure~\ref{fig:Sb2O3_2L}(c), exhibits no notable changes and is very similar to the interlayer charge density of the (Sb$_2$O$_3$)$_\parallel$ structure. The same is true for the electronic band structure in Figure~\ref{fig:Sb2O3_2L}(d), which features no major differences. However, the electronic band gap is slightly larger at a value of 4.6\,eV, which is closer to the electronic band gap of monolayer Sb$_2$O$_3$. Again, this suggests that the two layers in the (Sb$_2$O$_3$)$_\perp$ stacking configuration are more decoupled from each other. This assumption is further supported by the fact that even shifting one layer with respect to the other is not causing the layers to relax back into their original positions, and the resulting charge density and electronic band structure are almost identical to those found for the non-shifted (Sb$_2$O$_3$)$_\perp$ configuration. The interlayer charge density and the electronic band structure for the shifted (Sb$_2$O$_3$)$_\perp$ structure are depicted in the SI in Figure~S4.

\subsection{Twisted Heterobilayer Structures}\label{sec:SbS2O3}

\begin{figure*}[t]
    \centering
    \includegraphics[width=\linewidth]{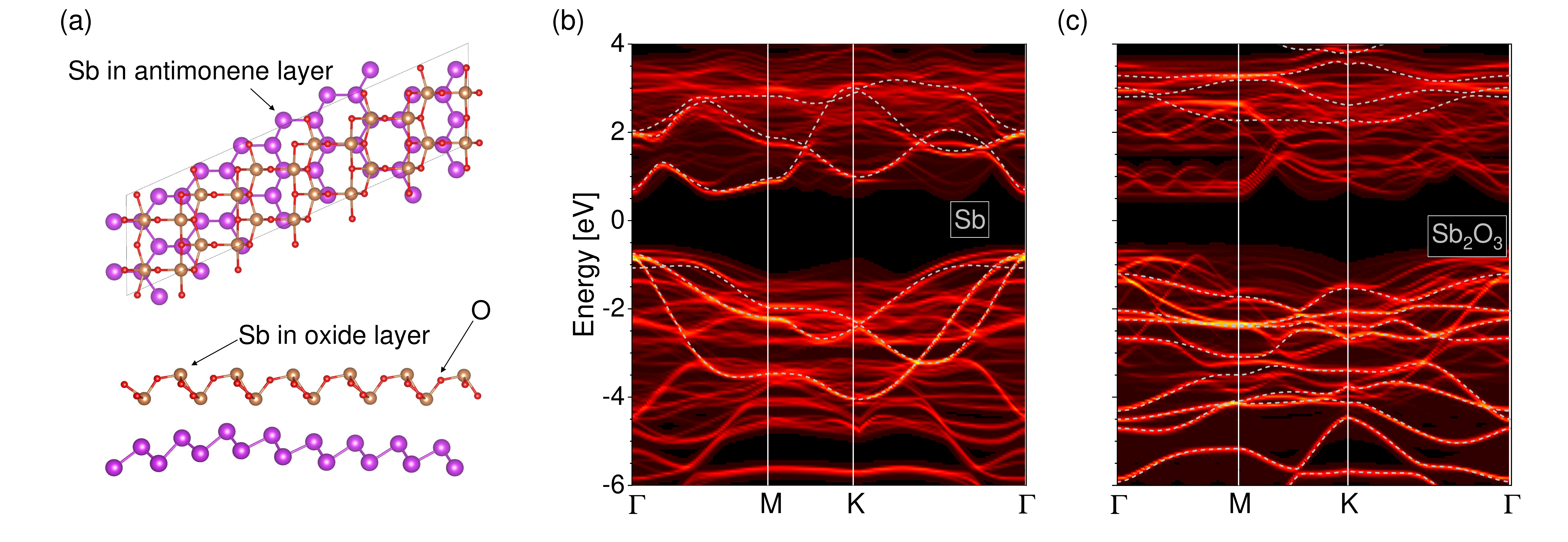}   
    \caption{Top and side view of a heterobilayer unit cell comprised of a pristine antimonene and Sb$_2$O$_3$ antimony oxide layer (a). The Sb and Sb$_2$O$_3$ sublayers were initially stretched and compressed by 1.5\,\%, respectively, enabling the creation of a coincidental lattice. Antimony atoms of the antimonene and antimony oxide sublayers are colored differently for easy distinction. Effective bandstructures (EBS) of the Sb and Sb$_2$O$_3$ sublayers obtained from unfolding of the electronic structure of the supercell (on the PBE level of theory) unfolded onto the primitive unit cells of antimonene and antimonene oxide are shown in (b) and (c), respectively. The bands were shifted such that the zero of energy coincides with the Fermi energy of the heterostructure. Dashed lines show the band structures of the individual monolayers shifted to coincide with the corresponding bands in the supercell for illustrative purposes. The EBS reveal significant hybridization between the two layers. The contributions of the Sb and Sb$2$O$3$ sublayers to the density of states support the picture of a pseudo-type-I band alignment in the heterostructure (d).}
    \label{fig:hetero_bilayer}
\end{figure*}

To further investigate the properties of the interlayer interactions of antimonene and oxidized antimonene, we created a heterobilayer consisting of a pristine antimonene layer and an Sb$_2$O$_3$ antimony oxide layer. The construction of a suitable coincidence supercell that possesses the correct periodicity is complicated compared to the case of homobilayers due to the different lengths and angles of the lattice vectors in the two materials. Additionally, in contrast to typical van-der-Waals-bound heterobilayers, e.g. involving transition-metal dichalcogenides, the more covalent interlayer interaction between antimonene (and also antimony oxide) layers induces a non-negligible change in the sublayer lattice constants in the heterostructure compared to the isolated monolayer materials. The construction of a strain-free heterobilayer hence requires careful choice of the coincidence lattice taking interlayer interactions into account.

To obtain a first assessment of the effect of interlayer interactions on the electronic structure of Sb/Sb$_2$O$_3$ heterobilayers, we hence opted to simulate a rather simple and small system by constructing two material layers: An Sb$_2$O$_3$ sublayer with the lattice vectors $\bm{a}_\text{Sb$_2$O$_3$}^{sc}=6\bm{a}_\text{Sb$_2$O$_3$}$, $\bm{b}_\text{Sb$_2$O$_3$}^{sc}=2\bm{b}_\text{Sb$_2$O$_3$}$, and an Sb sublayer with the lattice vectors $\bm{a}_\text{Sb}^{sc}=-\bm{a}_\text{Sb}-8\bm{b}_\text{Sb}$, $\bm{b}_\text{Sb}^{sc}=-2\bm{a}_\text{Sb}$. Here, $\bm{a}_\text{Sb}$ and $\bm{b}_\text{Sb}$, and $\bm{a}_\text{Sb$_2$O$_3$}$ and $\bm{b}_\text{Sb$_2$O$_3$}$ are the primitive cell lattice vectors of the antimonene and Sb$_2$O$_3$ layer, respectively. The lattice vectors of the supercells are indicated with the superscript $sc$. The two sublayers contain a lattice mismatch of about 3\,\% and a small difference in angle between the respective in-plane lattice vectors. We then combined the two sublayers into a single supercell by straining and slightly distorting the sublayers and optimized the atomic positions and in-plane lattice vectors to obtain the structural configuration with the smallest interatomic forces. The resulting supercell is depicted in Fig.~\ref{fig:hetero_bilayer} (a).

For a direct insight into the effect of hybridization on the electronic structure, we computed the effective bandstructures (EBS) of the Sb and Sb$_2$O$_3$ sublayers, shown in Fig.~\ref{fig:hetero_bilayer} (b) and (c). The EBS are obtained by unfolding the electronic structure of the supercell onto the entire reciprocal space with subsequent refolding onto the primitive cells of the sublayers by computing a spectral function that filters out contributions incompatible with the translation symmetry of the isolated sublayer lattice.~\cite{unfolding,unfolding-bandup} Our simulations indicate a noticeable hybridization between the layers; all electronic bands in the considered energy window around the global Fermi energy contain contributions from both layers to varying degrees. The bands form a \mbox{pseudo-type I} heterostructure, where the valence and conduction band edges predominantly contain contributions from the antimonene layer, while electronic bands with predominantly Sb$_2$O$_3$ nature appear deeper in the valence and conduction bands. This conclusion from the EBS is further supported by the contribution of the individual sublayers to the density-of-states of the heterostructure, which is depicted in Fig.~\ref{fig:hetero_bilayer} (d). We further analyzed the charge density distribution (see SI Figure~S7) and found indications for localized interlayer hybridization at locations, where the interlayer distance is smaller. As for pristine monolayer antimonene, the fundamental band gap of the system appears between the $\Gamma$ point and the conduction band valley in ${\Gamma}M$ direction, the energy difference between this indirect band gap and the direct band gap at $\Gamma$ (0.1\,eV) is reduced compared to freestanding unstrained monolayer antimonene (0.3\,eV in our simulations) due to the strain imposed onto the antimonene sublayer. Comparing the unfolded bands to the bandstructure of a correspondingly strained and distorted isolated antimonene monolayer [depicted as dashed lines in Fig.~\ref{fig:hetero_bilayer} (b), (c)], we also find small deviations in the band dispersion, particularly for the valence and conduction bands closest to the Fermi energy. We expect that these effects are largely reduced for supercells with smaller imposed strain.

\section{Conclusion}
We have presented a comprehensive study of the electronic properties of 2D bilayer antimony oxide and heterobilayer structures of antimonene and antimony oxide utilizing DFT calculations. The exploration of the charge density evolution and electronic band structure of different configurations of these materials has shown how the degree of oxidation and the interlayer interactions can tune the electronic band gap. Here the extent of the charge density overlap serves as a valuable indicator of interlayer interactions. In bilayer antimonene, the $d$ orbitals of antimony atoms in neighboring layers lead to a significant overlap of interlayer charge density, resulting in strong interlayer interactions and in a narrow electronic band gap. Introducing oxygen reduces available orbitals for these interactions, leading to a noticeable decrease in charge density between the layers and an increase of the electronic band gap.

The investigation of the heterobilayer has shown that significant hybridization between the layers occurs, i.e. every electronic band within the examined energy range around the Fermi energy incorporates input from both layers to different extents. The antimonene layer predominantly is responsible for the valence and conduction band edges, whereas higher contributions from the Sb$_2$O$_3$ are found deeper in the conduction or valence band.

Understanding how the properties of antimonene-based 2D structures can be modified through controlled oxidation and stacking opens up new avenues for the design and engineering of 2D materials with tailored functionalities.

\section{Methods}
The DFT calculations were carried out using the \text{Quantum ESPRESSO} suite.~\cite{QE-2009,QE-2017} For the structural relaxation ultrasoft pseudopotentials sourced from the GBRV library~\cite{GBRV} were used. A generalized gradient approximation (GGA) utilizing the PBE exchange-correlation functional~\cite{pbe} was used and the structures were optimized until the forces on every atom are below 0.005\,eV\AA$^{-1}$. A converged plane wave energy cutoff of 60\,Ry (816.34\,eV) and a Monkhorst-Pack $k$-point grid of $14 \times 14 \times 1$ points was used for antimonene homobilayers. A $2 \times 5 \times 1$ Monkhorst-Pack grid was used for Brillouin zone integration in case of the Sb/Sb$_2$O$_3$ heterobilayer in Section~\ref{sec:SbS2O3}. The sizes of the unit cells in the direction perpendicular to the layers were set to about 30\,\AA\, to ensure that there are no interactions between neighboring cells.

In order to highlight the charge density that arises from the creation of a bilayer, the charge density of the separate isolated layers at frozen atomic positions has been subtracted from the bilayer charge density. For the evaluation of all charge densities, the charge density minus a superposition of atomic densities is plotted at an isovalue of $-0.0004$, using the XCrySDen visualization program.~\cite{xcrysden} The charge densities without the subtraction of the isolated monolayer densities can be seen in the SI in Figure~S5; they are plotted at an isovalue of $-0.005$.

An investigation of the spin orbit interactions in the structures was carried out by utilizing fully relativistic norm-conserving pseudopotentials from the Pseudo-Dojo database.~\cite{pseudodojo} The inclusion of spin orbit interactions has had virtually no effect on the electronic band structure. An exemplary comparison is shown for the (Sb$_2$O$_2$)$_\parallel$ in the SI in Figure~S6.

Since the electronic band gap is generally underestimated by common DFT approximations, the electronic band structures in Section~\ref{sec:SbSb} on were calculated using the HSE12 hybrid functional.~\cite{hse12} For these calculations the structures were first relaxed using norm-conserving pseudopotentials from the SPMS repository~\cite{SPMS2023} and an energy cutoff of 80\,Ry (1088.46\,eV), which were also applied for the band structure calculations. All pseudopotentials utilize 6 and 15 valence electrons for oxygen and antimony respectively. Localized Wannier functions as implemented in the \texttt{wannier90} software package~\cite{wannier90} were applied to extract the electronic HSE12 band structures. A $9 \times 9 \times 1$ Monkhorst-Pack $k$-point grid was employed.

For the heterobilayer in Section~\ref{sec:SbS2O3}, we unfolded the electronic bandstructure of the supercell to the unitcells of the individual sublayers using the BandUP code.~\cite{unfolding-bandup,unfolding-banduppy} To obtain the spectral plots for the unfolded bands shown in  Figure~\ref{fig:hetero_bilayer}, we broadened each band by a Gaussian with a standard deviation of 30\,meV.

\medskip
\begin{acknowledgments}
The authors gratefully acknowledge the scientific support and HPC resources provided by the Erlangen National High Performance Computing Center (NHR@FAU) of the Friedrich-Alexander-Universität Erlangen-Nürnberg (FAU) under the NHR project b181dc. NHR funding is provided by federal and Bavarian state authorities. NHR@FAU hardware is partially funded by the German Research Foundation (DFG) – 440719683. This work has been supported by the Deutsche Forschungsgemeinschaft (DFG) - 182849149 - within the CRC 953 (B13).
\end{acknowledgments}

\medskip
\textbf{Supporting Information} \par
Supporting Information is available from the author.

\medskip
\textbf{Conflict of Interest} \par
The authors declare no conflict of interest.

\medskip
\textbf{Data Availability Statement} \par
Research data are not shared.

\medskip

\end{document}